\documentclass[aps,prb,twocolumn,showpacs]{revtex4}

\usepackage{graphicx}% Include figure files
\usepackage{dcolumn}% Align table columns on decimal point
\usepackage{bm}% bold math

\begin{document}

\title{Theory of electrical spin-detection at a ferromagnet/semiconductor interface}

\author{Athanasios N. Chantis and Darryl L. Smith}
\affiliation{Los Alamos National Laboratory, Los Alamos, New Mexico, USA}

\date{\today}

\begin{abstract}
We present a theoretical
model that describes electrical spin-detection at a ferromagnet/semiconductor interface.
We show that the sensitivity of the spin detector has \emph{strong} bias dependence which,
in the general case, is \emph{dramatically different} from that of the tunneling current
spin polarization. We show that this bias dependence originates
from two distinct physical mechanisms:
1) the bias dependence of tunneling current spin polarization, which is of \emph{microscopic}
origin and depends on the specific properties of the
interface, and 2) the macroscopic electron spin transport properties in the
semiconductor. Numerical results show that the magnitude of
the voltage signal can be tuned over a wide range from the second effect
which suggests a universal method for enhancing electrical spin-detection
sensitivity in ferromagnet/semiconductor tunnel contacts.
Using first-principles calculations we examine the particular case
of a Fe/GaAs Schottky tunnel barrier and find very good agreement with experiment. We also predict the
bias dependence of the voltage signal for a Fe/MgO/GaAs tunnel structure spin detector.

\end{abstract}

\maketitle

\section{INTRODUCTION}
Semiconductor spintronics aims to harness the electron's spin degree of freedom in data storage and
processing, typically by utilizing heterostructures composed of a combination of
magnetic and non-magnetic materials \cite{zutic}.
A fundamental problem in semiconductor spintronics was to find ways to
electrically generate non-equilibrium
electron spin distributions in conventional semiconductors.
Efficient electrical spin injection from ferromagnetic contacts into semiconductors using spin dependent tunneling
~\cite{rashba00}, ~\cite{darryl01} was shown to overcome the 'conductivity mismatch problem' associated with
highly conductive metallic contacts \cite{Schmidt}. 
Spin polarization of the tunneling current originates from the spin dependence
of the electron wavefunctions and the densities of states of the ferromagnetic contact. 
The spin dependent tunneling approach was realized experimentally using Fe interfaces with
GaAs ~\cite{hanbicki02,hanbicki03,crowell05,crooker,louNP},
silicon~\cite{huang,vantErve} and graphene~\cite{tombros}.
Jiang \emph{et. al} using CoFe/MgO interfaces showed enhanced
electrical spin injection efficiency into GaAs\cite{jiang}. Contacts made of CoFe
were also used to inject directly into GaAs~\cite{kotissek}
and showed an electron polarization that had dramatically different
bias dependence from that of Fe/GaAs contacts~\cite{louNP}.
In addition to electrical spin injection, efficient electrical spin detection
is required to achieve functional semiconductor spintronic devices. 

Crooker \emph{et al}~\cite{expt} recently reported experiments
of electrical spin detection using Fe/GaAs Schottky tunnel barriers as electrical spin detectors.
They demonstrated that both the magnitude and sign of the spin detection
sensitivity are tunable with voltage bias applied across the Fe/GaAs
interface; in some cases they were able to improve the spin detection sensitivity
by an \emph{order of magnitude}. The bias dependence of
the sensitivity of the detector was shown to be \emph{dramatically different} from that of the injected current
spin polarization.  A theoretical model was used to correlate
the spin-detection sensitivity of the Fe/GaAs
electrodes with their bias-dependent spin injection properties.
The model described successfully many of the experimentally observed trends~\cite{expt}.

Here we give a detailed description of this theoretical model of electrical
spin detection at a ferromagnet/semiconductor interface.
We consider a case when spin polarization is generated in the semiconductor by an external source
(e.g., optical or electrical) and subsequently detected at a ferromagnetic contact in which a tunnel barrier exists at the ferromagnet/semiconductor interface.
We incorporate first principles calculations to examine two
specific cases of tunnel barrier, a Fe/GaAs Schottky tunnel barrier and a Fe/MgO/GaAs tunnel structure. 
In this way we demonstrate that
the theory is general and can be applied to a variety of electrical spin detectors.
We show that the sensitivity of the electrical spin detectors
have strong bias dependence of both \emph{microscopic} and \emph{macroscopic} origin.
While the bias dependence of \emph{microscopic} origin is specific to each ferromagnet/semiconductor
tunneling structure, the \emph{macroscopic} bias dependence is general and depends on
the electrical transport properties of the semiconductor.

This article is organized as follows. In Section II we give a
detailed description of the theory. At the end of the section we
provide some approximate analytical formulas and look at the asymptotic limit
of large currents. This helps in understanding the general trends predicted
by the theory and facilitates a discussion of the numerical results. In section
III, first we present numerical results for a generic spin detector with a tunneling 
current polarization that has no bias dependence, then we incorporate
first principles calculations to examine the specific cases of Fe/GaAs Schottky barrier and Fe/MgO/GaAs
tunnel structure spin detectors. We conclude the paper in Section IV. with
a summary of our results and conclusions.  Some calculational details are included in an appendix.

\section{THEORETICAL MODEL}

A voltage signal, in response to a change in the spin polarization of the current, at a
ferromagnetic tunnel junction results because the tunneling resistance of the junction
depends on electron spin.  For example, if the tunneling resistance of junction is
smaller for spin-up (majority spin in the ferromagnetic contact) electrons than for
spin-down electrons (minority spin in the ferromagnetic contact), the voltage drop across the junction
will decrease if the spin-up electron component of the current increases
while the spin down electron component of the current decreases keeping the total current constant.
In recent experiments reported by Crooker et al~\cite{expt}, the spin polarization of the current at the
tunnel junction is changed either
by the absorption of circularly polarized light in the semiconductor in the vicinity
of the ferromagnetic tunnel junction or by electrical injection from a remote ferromagnetic contact.
The spin polarized electron density generated by the absorption of the circularly
polarized light or remote electrical injection drift/diffuses to the detection ferromagnetic tunnel junction and thus modifies the spin
polarization of the current across this junction.  A bias dependence of the voltage signal,
as is observed experimentally, occurs because of a combination of two effects:
1) transport of the remotely generated spin polarized
electron density depend on bias so that the change in spin polarization of the current crossing
the tunnel junction depends on voltage bias; and 2) the spin dependence of the tunneling
resistance depends on voltage bias.

\begin{figure}[tpb]

\includegraphics[width=.35\textwidth]{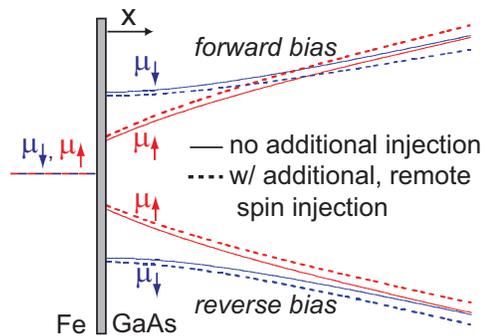}

\caption{ \small Schematic of
$\mu_{\uparrow,\downarrow}$ near a Fe/GaAs Schottky detector, at forward and
reverse bias.  Dotted lines show how $\mu_{\uparrow, \downarrow}$ are
modified by an additional, remote source of spin polarization.
In reverse bias remote spin generation increases
$ \left( \mu_ \uparrow - \mu_ \downarrow \right) $ near the contact but not
as much as in the region of spin generation.
}
\label{fig:Fig0}
\end{figure}

Fig.~\ref{fig:Fig0} shows
a schematic diagram of the electro-chemical potentials for spin-up and spin-down
electrons in the vicinity of a ferromagnetic tunnel junction in which the junction resistance
is smaller for spin-up electrons than for spin-down electrons. We show for the cases
under reverse bias (electron injection into the GaAs) and under forward bias 
(electron extraction from the GaAs), without remote spin generation (solid lines) and 
with remote generation of spin-up electrons (dashed lines).
The Fe contact is highly conductive so that the electric
field in the Fe is very small and the spin-up and spin-down electron distributions
in the Fe are very nearly in equilibrium with each other.  Thus, in the Fe, the electro-chemical
potentials for the two spin types are nearly constant in position and equal to each other
independent of bias.  The shaded stripe in Fig.\ref{fig:Fig0} represents the depletion
region in the GaAs that forms the tunnel barrier.  There is a drop in the
electro-chemical potential across this tunnel barrier region for both spin-up and spin-down
electrons.  This drop in electro-chemical potential is proportional to the product of
the current and the tunnel resistance for each spin type.  It is larger for spin-down electrons
than for spin-up electrons because the tunneling resistance is larger for
spin-down electrons than for spin-up electrons.  In reverse bias, the electro-chemical
potentials are higher in the Fe contact than in the GaAs and because the drop across the
junction is larger for spin-down electrons than for spin-up electrons there is a
surplus of spin-up electrons compared to spin down electrons near the junction interface.
In forward bias the electro-chemical potentials are higher in the GaAs than is the Fe
contact and there is a surplus of spin-down electrons compared to spin-up electrons
near the junction interface.  Because of spin relaxation, the electro-chemical potentials
for spin-up and spin-down electrons come together far from the tunnel junction when there is
no spin generation.  When there is generation of spin-up electrons,
the two electro-chemical potentials far from the tunnel junction remain
separated and are determined by the spin generation profile.  At any point in space
the spin polarization of the density is determined by statistics from the electro-chemical
potentials and the spin polarization of the current is determined by the product of the density and
the spatial derivative of the electro-chemical potential for each spin type.
The voltage signal at the tunnel junction produced by optical spin generation depends on
the change in spin polarization of the current at the tunnel junction and the tunneling resistance of
each spin type.

To describe theoretically the voltage signal we use the 1-dimensional model of
Ref.~\onlinecite{darryl01} with some modifications.  In this model 
the current flow at the interface is described using a spin dependent interface conductance:
\begin{equation}
j^{0}_{\eta}=G_{\eta}(\Delta \mu_{\eta}/e)
\label{eq:interface}
\end{equation}
where $j^{0}_{\eta}$ is the current density at the interface, $G_{\eta}$ is
the interface conductance, $\Delta \mu_{\eta}$ the interfacial
discontinuity in electrochemical potential for electrons with spin projection $\eta$, and
$e$ is the magnitude of the electron charge.
We set the interface at the position $x=0$ with the ferromagnet on the left side ($x<0$) and the
semiconductor on the right ($x>0$). As in Ref.~\onlinecite{darryl01}, we define
a variable $\beta$ such that $j_{\uparrow}=\beta j$, where $j$ is the total current density (independent of position in
the 1-dimensional model) and
a variable $\alpha$ such that $n_{\uparrow}=\alpha n$, where $n=n_\uparrow + n_\downarrow$ is the electron density
in the semiconductor which is independent of position and of bias.
Eq. (\ref{eq:interface}) can be written as:

\begin{equation}
(\mu^{R}_\uparrow(0)-\mu^{L}_\uparrow(0))=\frac{e j \beta^{0}}{G_\uparrow}
\label{eq:bcup}
\end{equation}
and

\begin{equation}
(\mu^{R}_\downarrow(0)-\mu^{L}_\downarrow(0))=\frac{e j (1-\beta^{0})}{G_\downarrow},
\label{eq:bcdown}
\end{equation}
where $\mu^{R,L}_\eta(0)$ is the electro-chemical potential for electrons at the right (left) side of
the interface and $\beta^0$ is $\beta$ evaluated at the interface.  Adding equations (\ref{eq:bcup}) and (\ref{eq:bcdown}) gives

\begin{eqnarray}
\left(\mu^{R}_{\uparrow}(0)+\mu^{R}_{\downarrow}(0)\right)=
\left(\mu^{L}_{\uparrow}(0)+\mu^{L}_{\downarrow}(0)\right)+\nonumber\\
 e j \left( \beta^{0} \left( \frac{1}{G_{\uparrow}}-\frac{1}{G_{\downarrow}} \right) +
\frac{1}{G_{\downarrow}} \right)
\label{eq:eqn0}
\end{eqnarray}
and subtracting them yields

\begin{eqnarray}
\left(\mu^{R}_{\uparrow}(0)-\mu^{R}_{\downarrow}(0)\right)=\left(\mu^{L}_{\uparrow}(0)-\mu^{L}_{\downarrow}(0)\right)+\nonumber\\
e j \left( \beta^{0} \left( \frac{1}{G_{\uparrow}}+\frac{1}{G_{\downarrow}} \right) -
\frac{1}{G_{\downarrow}} \right).
\label{eq:eqn1}
\end{eqnarray}
The voltage drop at the interface is

\begin{equation}
V=\left[
\left(\mu^{R}_{\uparrow}(0)+\mu^{R}_{\downarrow}(0)\right)
-\left(\mu^{L}_{\uparrow}(0)+\mu^{L}_{\downarrow}(0)\right)\right]/2e
\label{eq:voltage1}
\end{equation}
so that

\begin{equation}
V= (j/2) \left( \beta^{0} \left( \frac{1}{G_{\uparrow}}-\frac{1}{G_{\downarrow}} \right) +
\frac{1}{G_{\downarrow}} \right) .
\label{eq:voltage2}
\end{equation}

We consider a case in which the electrons in the semiconductor are strongly degenerate and
relate the electrochemical potentials and the densities using a zero temperature Fermi function, then

\begin{equation}
\mu^{R}_{\uparrow}(0)=2^{2/3}E_{F} \left( \alpha^{0} \right)^{2/3}
\label{eq:chemical1}
\end{equation}
and

\begin{equation}
\mu^{R}_{\downarrow}(0)=2^{2/3}E_{F}
\left( 1 - \alpha^{0} \right)^{2/3}
\label{eq:chemical2}
\end{equation}
where $E_{F}$ is the Fermi energy and $\alpha^0$ is $\alpha$ evaluated at the interface.

We describe the current flow in the semiconductor using spin dependent
drift-diffusion equations

\begin{equation}
j_{\eta}= e D \frac{\partial n_{\eta}}{\partial x} + e \mu E n_{\eta}
\label{eq:drift-diff}
\end{equation}
where $D$ is the diffusion coefficient, $E$ is the electric field in the semiconductor (independent of position in
the 1-dimensional model), and
$\mu$ is the electron mobility.  The drift-diffusion equation at the interface gives:

\begin{equation}
j^{0}_{\uparrow}-j^{0}_{\downarrow}= (2 \beta^{0} -1)j = e D \frac{\partial \Delta n}{\partial x}|_{x=0} + e \mu E \Delta n(0)
\label{eq:eqn2}
\end{equation}
If electrons with different spins are driven out of local quasi-thermal
equilibrium at some region in space, so that $n_{\uparrow}$ is not equal to $n_{\downarrow}$,
the difference in the two electron densities $\Delta n(x)=(n_{\uparrow}-n_{\downarrow})(x)$ relax
as described by the spin current continuity equation

\begin{equation}
D \frac{\partial^{2}\Delta n}{\partial x^{2}} + \mu E \frac{\partial \Delta n}{\partial x}=\frac{\Delta n}{\tau_s}
-f(x)
\label{eq:spin-diff}
\end{equation}
where $\tau_s$ is the spin-relaxation time and $f(x)$ is the spin generation function.

We are interested in how the interface voltage drop varies with small changes in the amplitude $C$ of the spin generation
function $f(x)=(C/\tau_s)F(x)$ where
$F(x)$ is a unitless and normalized function of position. The voltage signal depends on the amplitude of the
spin generation function through the spin polarization of the current at the interface Ref.~\onlinecite{G(V)}
\begin{equation}
\frac{dV}{dC}|_{C=0}=(1/2)\left( \frac{1}{G_{\uparrow}}-\frac{1}{G_{\downarrow}} \right)j\frac{d\beta^{0}}{dC}|_{C=0}
\label{sds}
\end{equation}

The voltage signal depends on the derivative of
$\frac{\partial \Delta n}{\partial x}|_{x=0}$ and $\Delta n(0)$ with respect to C and to calculate these derivatives we use
Eq. (\ref{eq:spin-diff}). We search for a solution of this equation
of the form $\Delta n(x)= Q(x) + \int^{\infty}_0 f(x^\prime)g(x,x^\prime)dx^\prime$
subject to boundary conditions $\Delta n(0)=Q_0$ and $\Delta n(\infty)=0$
where $Q(x)=Q_0 e^{-x/\Lambda_+}$ is the solution of the homogeneous differential equation and
$g(x,x^\prime)$ is the Green's function solution of

\begin{equation}
D \frac{\partial^{2}g(x,x^\prime)}{\partial x^{2}} + \mu E \frac{\partial g(x,x^\prime)}{\partial x}-\frac{g(x,x^\prime)}{\tau_s}=-\delta\left(x-x^\prime\right)
\label{eq:green}
\end{equation}
with boundary conditions $g(0,x^\prime)=0$ and $g(\infty,x^\prime)=0$.
For a general spin generation function, we have

\begin{equation}
\frac{\partial \Delta n}{\partial x}|_{x=0}=-\frac{\Delta n(0)}{\Lambda_+}
+\int^{\infty}_0 f(x^\prime)\frac{\partial g(x,x^\prime)}{\partial x}|_{x=0}dx^\prime
\label{eq:eqn3}
\end{equation}
where

\begin{equation}
g(x,x^\prime) = \left\{ \begin{array}{rl}
-\frac{1}{D}
\frac{e^{\frac{x^\prime}{\Lambda_-}}
\left[e^{-\frac{x}{\Lambda_+}}-e^{-\frac{x}{\Lambda_-}}\right]}
{\left[\frac{1}{\Lambda_+}-\frac{1}{\Lambda_-}\right]}
 &\mbox{, $0\le x \le x^\prime$}\\
&\\
-\frac{1}{D}
\frac{e^{-\frac{x}{\Lambda_+}}
\left[e^{\frac{x^\prime}{\Lambda_-}}-e^{\frac{x^\prime}{\Lambda_+}}\right]}
{\left[\frac{1}{\Lambda_+}-\frac{1}{\Lambda_-}\right]}
&\mbox{, $x^\prime \le x$}
\end{array} \right.
\label{eq:g}
\end{equation}
and

\begin{equation}
\frac{1}{\Lambda_\pm(E)}=
\frac{1}{\sqrt{D \tau_s}}
\left[ \left( \frac{\mu E \tau_s}{2\sqrt{D \tau_s}} \right)
\pm
\left[\left(\frac{\mu E \tau_s}{2\sqrt{D \tau_s}}\right)^2
+1\right]^{\frac{1}{2}}
\right].
\label{eq:lambda}
\end{equation}
Using $j=\sigma E$, where $\sigma =e \mu n$ is the conductivity of the semiconductor, $\Lambda_s = \sqrt{D\tau_s}$ and the
Fermi liquid relationship $D/\mu=2E_F/3e$, we can write $\Lambda_\pm$ as

\begin{equation}
\frac{1}{\Lambda_\pm(j)}=
\frac{1}{\Lambda_s}
\left[ \left( \frac{3 e j \Lambda_s}{4\sigma E_F} \right)
\pm
\left[\left(\frac{3 e j \Lambda_s}{4\sigma E_F}\right)^2
+1\right]^{\frac{1}{2}}
\right]
\label{eq:lambda}
\end{equation}
To be specific, we consider a striped spin generation function of the form:

\begin{equation}
F(x) = \left\{ \begin{array}{rl}
1 &\mbox{if $y-d/2\le x \le y+d/2$}\\
0&\mbox{otherwise}
\end{array} \right.
\label{eq:g}
\end{equation}
where $y$ is the center position of striped spin generation and $d$ is its width ($y\ge d/2$).  Then Eq. (\ref{eq:eqn2}) can be written as

\begin{equation}
j^{0}_{\uparrow}-j^{0}_{\downarrow}=
-\frac{eD}{\Lambda_+}\left[\Delta n(0)+C F(j)
\right]+e\mu E \Delta n(0)
\label{eq:eqn4}
\end{equation}
where in general

\begin{equation}
F(j)=\int^{\infty}_0 F(x) e^{x/\Lambda_-}d(x/\Lambda_-).
\end{equation}
For the striped spin generation function $F(j)= e^{y/\Lambda_-}\left(e^{d/2\Lambda_-}-e^{-d/2\Lambda_-}\right)$ and
depends on current through $\Lambda_-$. (F(j) is negative because $\Lambda_-$ is negative).  Because
$j^{0}_{\uparrow}+j^{0}_{\downarrow}=e\mu E n$, $j^{0}_{\uparrow}-j^{0}_{\downarrow}=(2\beta^{0}-1)j$ and $\Delta n(0)=n(2\alpha^{0}-1)$
this can be written in terms of the values of parameters $\alpha$ and $\beta$ at the interface.

\begin{eqnarray}
\lefteqn{(2\beta^{0}-1)j=}\nonumber\\
&&-\frac{2 E_F \sigma}{3 e \Lambda_+}
\left[(2\alpha^{0}-1)+(C/n) F(j)\right]+\nonumber\\
&&(2\alpha^{0}-1)j
\label{eq:eqn5}
\end{eqnarray}

Considering that on the metal side of the interface the
electrochemical potentials for spin $\uparrow$ and $\downarrow$ electrons are
very nearly equal, Eqs. (\ref{eq:eqn1}), (\ref{eq:chemical1}) and (\ref{eq:chemical2}) give

\begin{eqnarray}
2^{\frac{2}{3}}E_F
\left[\left(\alpha^{0}\right)^{\frac{2}{3}}-\left(1-\alpha^{0}\right)^{\frac{2}{3}}\right]=\nonumber\\
ej\left[\beta^{0}\left(\frac{1}{G_\uparrow}+
\frac{1}{G_\downarrow}\right)-\frac{1}{G_\downarrow}\right]
\label{eq:eqn6}
\end{eqnarray}
Equations (\ref{eq:eqn5}) and (\ref{eq:eqn6}) together give $\beta^{0}$ and $\alpha^{0}$ for any given value of the amplitude $C$ of the source function and of bias.
Eq. (\ref{eq:eqn6}) is non-linear and is solved numerically for
$\alpha^{0}$ after elimination of $\beta^{0}$ with the help of Eq. (\ref{eq:eqn5}).
The spin-detection sensitivity, defined as $dV/C$, and the current polarization, $2\beta^0 -1$, are then calculated numerically.

Before we present the numerical results of the model, it is instructive to linearize
Eq. (\ref{eq:eqn6}), in order to obtain an analytic expression for $dV/dC$ valid for weak injection conditions.
For the linear case it is convenient to rewrite Eqs. (\ref{eq:eqn5}) and  (\ref{eq:eqn6})
in terms of a parameter $\delta^0=\alpha^0 - 1/2$ because $\alpha^0$ is close to $1/2$:

\begin{equation}
(2\beta^{0}-1)j=-\frac{4 E_F \sigma}{3 e \Lambda_+}
\left[\delta^{0}+\delta^{d}F(j)\right]+2\delta^{0}j
\label{eq:eqn5d}
\end{equation}

\begin{eqnarray}
2^{\frac{2}{3}}E_F
\left[\left(\frac{1}{2}+\delta^{0}\right)^{\frac{2}{3}}-\left(\frac{1}{2}-\delta^{0}\right)^{\frac{2}{3}}\right]=\nonumber\\
ej
\left[\beta^{0}\left(\frac{1}{G_\uparrow}+
\frac{1}{G_\downarrow}\right)-\frac{1}{G_\downarrow}\right]
\label{eq:eqn6d}
\end{eqnarray}
where $\delta^{d}= {C \over 2n}$. We
eliminate $\beta^{0}$ from the second equation and differentiate with respect to $\delta_d$

\begin{equation}
j\frac{d\beta^{0}}{d\delta^d}=-\frac{2 E_F \sigma}{3 e \Lambda_+}
\left[\frac{d\delta^{0}}{d\delta^d}+F(j)\right]+j\frac{d\delta^{0}}{d\delta^d}
\label{eq:eqn5d_diff}
\end{equation}

\begin{equation}
 \frac{d\delta^{0}}{d\delta^d}=-\frac{F(j)\frac{\sigma}{\Lambda_+}}
{A(j)}
\label{eq:eqn6d_diff}
\end{equation}
where

\begin{eqnarray}
A(j)=-\frac{\sigma}{\Lambda_-} + \nonumber\\
\frac{2^{\frac{2}{3}}\left[\left(\frac{1}{2}+\delta^{0}\right)^{-\frac{1}{3}}
+\left(\frac{1}{2}-\delta^{0}\right)^{-\frac{1}{3}}\right]}
{\left(\frac{1}{G_\uparrow}+\frac{1}{G_\downarrow}\right)}
\end{eqnarray}
When the density polarization is small $\delta^0 \to 0$
and $\left[\left(1/2+\delta^{0}\right)^{-1/3}
+\left(1/2-\delta^{0}\right)^{-1/3}\right] \approx 2^{4/3}$.
In this case, we can linearize $A(j)$

\begin{equation}
A(j)=-\frac{\sigma}{\Lambda_-} +
\frac{4}{\left(\frac{1}{G_\uparrow}+\frac{1}{G_\downarrow}\right)}.
\label{eq:linA}
\end{equation}
The current polarization is given by,

\begin{equation}
(2\beta^0 -1) =\frac{(\frac{\sigma}{\Lambda_-}) \left( \frac{\frac{1}{G_\uparrow}-\frac{1}{G_\downarrow}}
{\frac{1}{G_\uparrow}+\frac{1}{G_\downarrow}}\right)}{A(j)}.
\label{eq:linAA}
\end{equation}
We note in passing that equations (\ref{eq:linAA}) and (\ref{eq:voltage2}) combined
can be used to extract $G_\uparrow$ and $G_\downarrow$ from the exerimental
data set of total current, voltage and current polarization.

Equations (\ref{eq:linA}), (\ref{eq:eqn6d_diff}) and (\ref{eq:eqn5d_diff})
give an analytic expression for $dV/dC$, which in terms of the parameter $\delta^d$
can be written as

\begin{equation}
\frac{dV}{dC}|_{C=0}=(1/2n)\frac{dV}{d\delta^d}|_{\delta^d=0}
=(1/2n)\left(\frac{1}{G_\uparrow}-\frac{1}{G_\downarrow}\right)
j\frac{d\beta^{0}}{d\delta^d}
\end{equation}

In the derivation of Eq. (\ref{eq:eqn5d_diff}) we kept the diffusion (first RHS
term) and drift (second RHS term) contributions to the voltage change separate. The diffusion term has two contributions.
The term $-(2E_F\sigma/3e\Lambda_+)F(j)$ is always positive.  It describes the change of
current polarization at the interface, due to diffusion, resulting from the Green's function term in Eq. (\ref{eq:eqn3}). The term
$-(2E_F\sigma/3e\Lambda_+)\frac{d\delta^{0}}{d\delta_d}$ is always negative because $\frac{d\delta^{0}}{d\delta^d}$ is positive.  It describes the change of
current polarization at the interface, due to diffusion, resulting from the homogeneous solution term in Eq. (\ref{eq:eqn3})
The drift term, $ j\frac{d\delta^{0}}{d\delta_d}$, describes the change in spin current polarization at the contact from drift of the
modified spin density polarization at the interface due to external spin generation.
Its sign depends on the sign of $j$ .  For \emph{spin injection} ($j<0$)
the absolute value of $\frac{d\delta^{0}}{d\delta_d}$
cannot be larger than $\left|F(j)\right|$ and hence the diffusion term is
always positive.  The drift term is always negative and therefore the
two processes always oppose each other for spin injection. For \emph{spin collection} ($j>0$) $\frac{d\delta^{0}}{d\delta_d}$
is smaller than $\left|F(j)\right|$ for small currents but can become larger than $\left|F(j)\right|$ for large currents.  As
a result the diffusion term is positive for small currents, but can become negative for large currents and
the drift and diffusion augment each other for small currents but can oppose each other for
large currents in spin collection.

It is interesting to examine the asymptotic behavior for large bias ~\cite{largebias} of $dV/dC$.
When $3e\left|j\right|\Lambda_s/(4\sigma E_F ) \gg 1$, we can write
\begin{equation}
\frac{1}{\Lambda_+(j)} \approx \frac{1}{\Lambda_s}
\left[\frac{3ej\Lambda_s}{4\sigma E_F}
+ \frac{3e\left|j\right|\Lambda_s}{4\sigma E_F}
+ \frac{1}{2\frac{3e\left|j\right|\Lambda_s}{4\sigma E_F}}\right]
\label{eq:Lexpanded}
\end{equation}
For the case of \emph{spin collection}, where j is positive, this becomes
as

\begin{equation}
\frac{1}{\Lambda_+(j)} \approx \frac{3ej}{2\sigma E_F}.
\end{equation}
Then we have

\begin{equation}
A(j)\approx\frac{4}{R_+},
\end{equation}
where $R_+=\left( 1/G_\uparrow+1/G_\downarrow\right)$ and

\begin{equation}
 \frac{d\delta^{0}}{d\delta^d}=-\frac{F(j)R_+}
{4}\frac{3ej}{2E_F}
\end{equation}
so that

\begin{equation}
j\frac{d\beta^{0}}{d\delta_d}\approx -jF(j)\left(1-\frac{3e}{8 E_F} R_+ j\right)
- j F(j)\frac{3e}{8 E_F} R_+ j
\end{equation}
Part of the diffusion term (RHS first term)
cancels exactly the drift term (RHS second term) leaving a linear dependence on $j$,

\begin{equation}
\frac{dV}{dC}|_{C=0}\approx -(1/2n)\left(\frac{1}{G_\uparrow}-\frac{1}{G_\downarrow}\right)F(j)j
\label{eq:vd_accum}
\end{equation}
Note that F(j) is always negative and for the case of a homogeneous source $F(j)=-1$.

For the case of \emph{spin injection}, where j is negative, Eq. (\ref{eq:Lexpanded})
can be written as

\begin{equation}
\frac{1}{\Lambda_+(j)} \approx \frac{1}{\Lambda_s}\frac{2 \sigma E_F}{3e\left|j\right|\Lambda_s}
\end{equation}
Then,
\begin{equation}
A(j)\approx\frac{3 e \left|j\right|}{2 E_F}+\frac{4}{R_+}+\frac{2E_F\sigma^2}{3e\left|j\right|\Lambda_s^2}
\end{equation}
and

\begin{equation}
 \frac{d\delta^{0}}{d\delta^d}=-\frac{F(j)}
{A(j)}\frac{2E_F \sigma^2}{3e\Lambda_s^2 \left|j\right|}
\end{equation}
so that

\begin{eqnarray}
j\frac{d\beta^{0}}{d\delta^d}\approx \left[ \frac{F(j)}
{A(j)} \left( \frac{2E_F \sigma}{3e\Lambda_s} \right)^2 \frac{1}{\left|j\right|} \right] \times \nonumber\\
\left\{\frac{2 E_F \sigma^2}{3 e \Lambda_s^2 \left|j\right|}\right. \nonumber\\
-\left( \frac{3e\left|j\right|}{2E_F}+\frac{4}{R_+} + \frac{2 E_F \sigma^2}{3 e \Lambda_s^2 \left|j\right|} \right)\nonumber\\
\left.+  \frac{3e\left|j\right|}{2E_F}\right \}
\label{eq:dbinj}
\end{eqnarray}
The first two RHS terms are the diffusion and the third the drift contributions to
the change of current spin polarization at the interface due to external spin
generation. The leading order terms in $\left|j\right|$ have opposite signs and
cancel. Then to leading order in $\left|j\right|$,

\begin{equation}
j\frac{d\beta^{0}}{d\delta^d}\approx
- F(j)\left(\frac{2 E_F}{3 e}\right)^3\left(\frac{\sigma}{\Lambda_s}\right)^2 \frac{4}{R_+\left|j\right|^2}
\end{equation}
and

\begin{eqnarray}
\frac{dV}{dC}|_{C=0}\approx -(1/2n)\left(\frac{1}{G_\uparrow}-\frac{1}{G_\downarrow}\right) \nonumber\\
\times F(j)\left(\frac{2 E_F}{3 e}\right)^3
\left(\frac{\sigma}{\Lambda_s}\right)^2 \frac{4}{R_+\left|j\right|^2}
\label{eq:vd_inj}
\end{eqnarray}
Unlike spin collection
where the spin detection sensitivity grows linearly with $\left|j\right|$, during spin injection
spin sensitivity drops as $1/\left|j\right|^2$.   It is worth mentioning that the exact
cancellation of leading order terms in Eq. (\ref{eq:dbinj}) occurs strictly
in the 1D case, but there is no physical principle that demands exact cancellation
for a more complicated geometry. This may lead to a \emph{reversal} of the sign of the voltage signal with bias for spin injection.

\section{RESULTS}
\begin{figure}[tpb]

\includegraphics[]{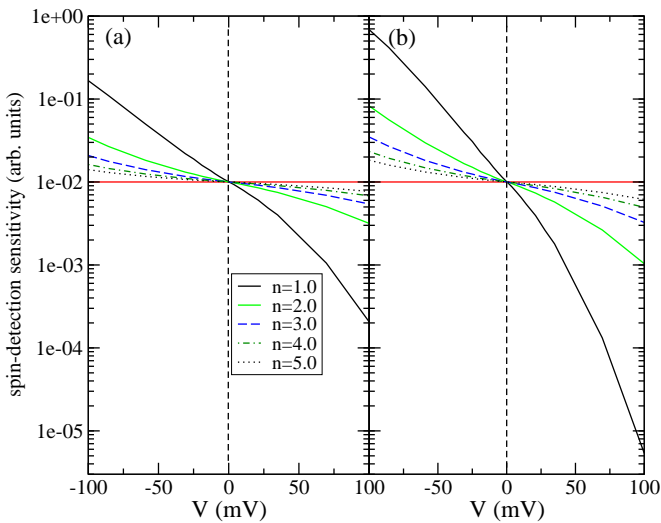}

\caption{ \small (Color online) The calculated spin detection sensitivity for an electrical spin detector
with various levels of doping and tunneling current spin polarization (solid red/gray line)
which is independent of bias. The vertical axis is logarithmic. The left panel
is for $\mu = 3000 cm^2V^{-1}s^{-1}$ and the right for $\mu = 1500 cm^2V^{-1}s^{-1}$.
The values of carrier concentrations used are depicted in the plot.
The spin-relaxation time, $\tau_s$, is set to $10^{-7} s$.
The source is set at a distance of
$2\times10^{-3} cm$ from the interface and it has a
width of $5 \times10^{-4} cm$. The source function is constant within this interval.
}
\label{fig:Fig1}
\end{figure}

\begin{figure}[tpb]

\includegraphics[]{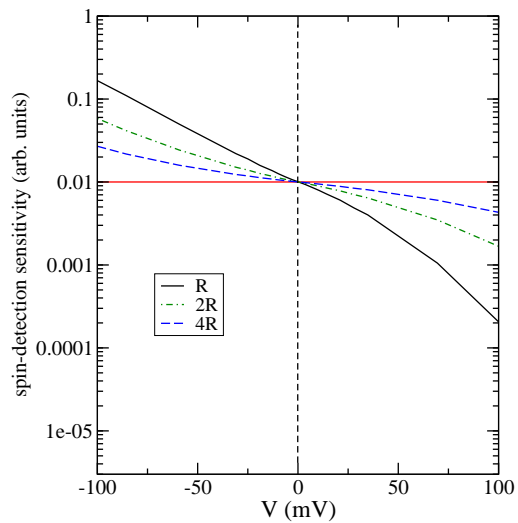}

\caption{ \small (Color online) The calculated spin detection sensitivity for an electrical spin detector
with tunneling current spin polarization which is independent of bias (solid red/gray line).
The vertical axis is logarithmic
For the case of $\mu = 3000 cm^2V^{-1}s^{-1}$ and $n=1\times10^{16} cm^{-3}$.
We vary the tunneling resistance by multiples of 2.
The spin-relaxation time is set $\tau_s=10^{-7} s$.
The source is set at a distance of
$2\times10^{-3} cm$ from the interface and it has a
width of $5 \times10^{-4} cm$. The source function is constant within this interval.
}
\label{fig:Fig2}

\end{figure}

We present results of the model from numerically solving the nonlinear equations. 
In the following, in order to facilitate the comparison with experiments we adopt the convension that
\emph{negative voltage} corresponds to \emph{spin collection} and \emph{positive voltage} to
\emph{spin injection}. Often, as is done in experimental works we will refer to the negative voltage
as \emph{forward bias} and to the positive voltage as \emph{reverse bias}.
In Fig.~\ref{fig:Fig1} we
show the calculated spin detection sensitivity
for an electrical spin
detector with an interface which has a constant current spin polarization
(independent of bias).
Required inputs to the calculation of the
voltage signal are: electron density, mobility, spin lifetime,
and the spin tunneling conductances $G_ {\uparrow, \downarrow}$.
The left panel in Fig.~\ref{fig:Fig1}
is for mobility $\mu =$ 3000 $cm^2V^{-1}s^{-1}$ and the right for $\mu =$ 1500 $cm^2V^{-1}s^{-1}$.
We have used several values for carrier concentrations and they are shown in the plot.
The spin-relaxation time is set to $\tau_s=10^{-7} s$.
The source is set at a distance of
$2\times10^{-3} cm$ from the Fe/GaAs interface and it has a
width of $5 \times10^{-4} cm$, the source function is constant within this interval.
As seen in Fig~\ref{fig:Fig1}, in all cases, for \emph{spin injection}
the magnitude of the calculated voltage signals
drops rapidly with increasing bias even though
the current polarization is constant.  By contrast,
the magnitude of the calculated voltage signals,
increases rapidly with increasing bias during \emph{spin collection}.
The difference between the calculated voltage signal and current polarization is larger for
smaller values of the electron mobility and electron concentration.
The magnitude of the voltage signal is
smaller than the current polarization in reverse bias but larger in forward bias
for a combination of two reasons: 1) the drift and diffusion contributions to Eq. (\ref{eq:eqn5d_diff})
oppose each other in spin injection but add in spin collection;
and 2) the electric field in the semiconductor tends to drift the optically or electrically
generated spin polarized electrons away from the detector contact in reverse bias but
toward the detector contact in forward bias.
When the semiconductor is more heavily doped the electric field is smaller therefore
these effects are less pronounced than when the semiconductor is more lightly doped.
More specifically, from equations (\ref{eq:vd_inj}) and (\ref{eq:vd_accum}) we can see that the
rate of drop/increase depends on the conducting properties of the semiconductor,
the spin polarization of the interface (prefactor $(1/G_\uparrow - 1/G_\downarrow)$) and the
location and width of the spin source (prefactor $F(j)$). The rate of increase during
spin collection is proportional to $\propto n^{-2/3}$ while the rate of decrease is
proportional to $\propto \mu^{-2} n^{-8/3}$

In Fig.~\ref{fig:Fig2} we show the influence of the tunneling resistance.
In a ferromagnet/semiconductor interface the tunneling resistance
depends on the Schottky barrier height, width and shape.
The first is determined by the magnitude of the band gap in the semiconductor and the position of the
Fermi level relative to the top of the valence band while the last two are modulated with doping.
In Fig.~\ref{fig:Fig2} we can see that smaller tunneling resistance results in bigger
variation of spin detection sensitivity with bias. This can be understood by from 
the linearized results in Eq. (\ref{eq:vd_inj}).

The rate of change of spin detection sensitivity with
bias is inversely proportional to the tunneling resistance. The physical origin
of this lies in that, for a given applied voltage,
larger tunneling resistance will result in smaller
current; this is reducing the effect of drift.
\begin{figure}[tpb]

\includegraphics[]{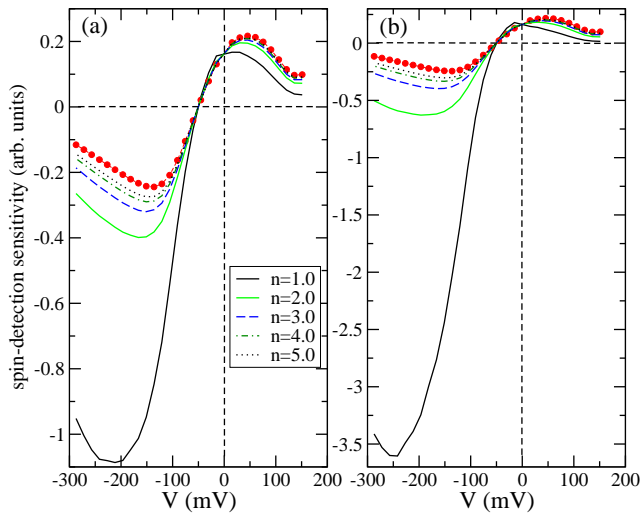}

\caption{ \small (Color online) The calculated spin detection sensitivity for an Fe/GaAs Schottky electrical spin detector
with various levels of doping.
On both panels, the red/gray solid line with circles is the Fe/GaAs tunneling current spin
polarization calculated from first-principles.
The left panel
is for $\mu = 3000 cm^2V^{-1}s^{-1}$ and the right for $\mu = 1500 cm^2V^{-1}s^{-1}$.
The values of carrier concentrations used are depicted in the plot.
The spin-relaxation time, $\tau_s$, is set to $10^{-7} s$.
The source is set at a distance of
$2\times10^{-3} cm$ from the Fe/GaAs interface and it has a 
width of $5 \times10^{-4} cm$. The source function is constant within this interval.
}
\label{fig:Fig3}
\end{figure}
Generally speaking, the current spin polarization and hence $(1/G_\uparrow-1/G_\downarrow)$ can
have a strong bias dependence. It was shown in Ref. \onlinecite{crooker,crowell,chantis} that
the current spin polarization of Fe/GaAs(001) junctions has a very strong bias
dependence, and even reverses sign within a small interval around zero bias.
In Ref.~\onlinecite{chantis,dery} two different microscopic models to explain the experimentally observed bias dependence of the tunneling current were discussed.
The bias dependence of spin sensitivity is due to a combination of 
the macroscopic physics described above and the microscopic bias dependence of
$(1/G_\uparrow-1/G_\downarrow)$.
To predict the resulting behavior in specific ferromagnet/semiconductor junctions
we have incorporated our model first-principle results for the bias dependence of $(1/G_\uparrow-1/G_\downarrow)(V)$.

\begin{figure}[tpb]

\includegraphics[]{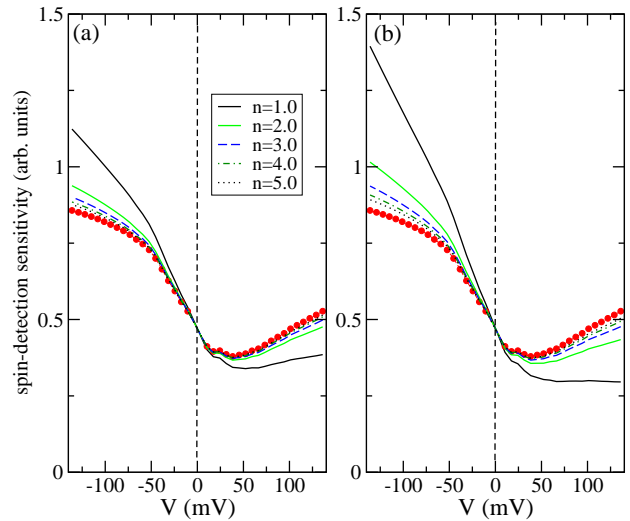}

\caption{ \small (Color online) The calculated spin detection sensitivity for an Fe/MgO/GaAs electrical spin detector
with various levels of doping.
On both panels, the red/gray solid line with circles is the Fe/MgO tunneling current spin
polarization calculated from first-principles. The left panel
is for $\mu = 3000 cm^2V^{-1}s^{-1}$ and the right for $\mu = 1500 cm^2V^{-1}s^{-1}$.
The values of carrier concentrations used are depicted in the plot.
The spin-relaxation time, $\tau_s$, is set to $10^{-7} s$.
The source is set at a distance of
$2\times10^{-3} cm$ from the Fe/GaAs interface and it has a 
width of $5 \times10^{-4} cm$. The source function is constant within this interval.
}
\label{fig:Fig4}
\end{figure}

In Fig.~\ref{fig:Fig3} we show the calculated spin detection sensitivity and first-principles current polarization
for a Fe/GaAs(001) interface. The first-principle method and results are identical to that presented in
Ref.~\onlinecite{chantis}.  As it was explained in Ref.~\onlinecite{chantis} the interface electronic structure of
Fe/GaAs(001) results in two strong minority-spin peaks in the energy dependence of
electron transmission across the interface. One is located at about 125 meV below $E_F$ and the other 125 meV above.
The result is a strong bias dependence of the current spin polarization. We see in Fig.~\ref{fig:Fig2} that the
bias dependence of spin detection sensitivity bares some resemblance to that of the current spin polarization but in the general case can 
be significantly different.
The magnitude of $dV/dC$ decreases (increases) faster than the spin polarization in the negative (positive) bias and
the difference between the two increases as we make the semiconductor \emph{less} conductive.
As we can see in Fig.~\ref{fig:Fig3} the spin detection sensitivity can be \emph{raised by an order of magnitude}
in the positive bias. Therefore, under certain conditions the macroscopic factors described above can
have a dominant influence over the microscopic factors that influence the bias dependence of current polarization.
Since it is much easier to control the conducting properties of semiconductor rather the microscopic
electronic properties of the interface, these effect can have a direct application in optimization of electrical spin detectors. This result is in very good agreement with the experimental bias dependence of spin-detection
sensitivity presented in Ref.~\onlinecite{expt}.

It is interesting to examine the bias dependence of spin-detection sensitivity for a  
different interface. The Fe/MgO(001) interface is different from Fe/GaAs(001) in many ways
and it is of great interest to spintronics community.    
To calculate the spin-dependent tunneling conductance of Fe/MgO(001) we used the same approach as in 
Ref.~\onlinecite{chantis,chantis2,chantis3} and the same setup with Ref.~\onlinecite{kirill}. 
The details of the calculation such as the chosen k-mesh in the two dimensional Brillouin zone (2DBZ)
and the method of calculation for the total current and spin polarization are the same with those
used for Fe/GaAs(001) interface and are described in Refs.~\onlinecite{chantis,chantis2}.
It was shown in Ref.~\onlinecite{kirill,khan}
that the interface minority-spin resonances in Fe/MgO(001) interface are located far from the $\Gamma$
point in the 2DBZ, contributing much less to the tunneling conductance than they do in the case
of Fe/GaAs(001) interface. Therefore as we can see in Fig.~\ref{fig:Fig4} the spin polarization
of the tunneling current has less dramatic bias dependence in this case.
The band gap of MgO is about 5 times larger than
the band gap of GaAs so the Fe/MgO tunneling barrier has much bigger resistance than the Fe/GaAs.
Because of these differences with Fe/GaAs interface the spin-detection sensitivity is
much less sensitive to the changes in applied bias. 
This is consistent with the analysis given so far and in particular with Eq.~(\ref{eq:vd_inj}).
\section{CONCLUSION}

We presented a theory of electrical spin detection in ferromagnet/semiconductor detectors.
We showed that the sensitivity of such detectors can have a \emph{strong} bias dependence. The origin
of this dependence lies in the microscopic electronic structure of the interface and the
macroscopic electrical properties of the conducting channel in the semiconductor.
The first was incorporated in our model with the help of first principles electronic structure
calculations.
With the help of a model spin detector which has constant current polarization with respect to bias
we showed that the latter by itself is capable of producing \emph{strong} bias dependence of
sensitivity. This result suggests that enhancement of detector's sensitivity is possible independent of
the materials used to construct the detector by engineering the electrical properties
of the conducting channel in the semiconductor and tuning the bias.
Our results for the particular case of Fe/GaAs Schottky tunnel contacts show a very good agreement with
experiment~\cite{expt}. As in the experiment we were able to enhance the spin sensitivity
by an order of magnitude when applied positive voltage. Our results for Fe/MgO/GaAs
show a similar enhancement though the magnitude of the effect is smaller than
in Fe/GaAs. This is explained by the bigger height of tunneling barrier in the
case of Fe/GaAs. These results suggest specified routes on how to engineer
efficient electrical spin detectors using ferromagnet/semiconductor interfaces.

\begin{acknowledgments}
This work was supported by DOE Office of Basic Energy Sciences Work Proposal Number 08SCPE973.  We thank S. A. Crooker and P. A. Crowell for many valuable discussions.

\end{acknowledgments}

\section{Appendix: Voltage Derivative of the Interface Conductances}

In the previous sections we neglected terms proportional
to the voltage derivative of the interface conductances because these
terms are small for typical parameter values.  In this appendix we discuss the contribution
of these terms.  For notational simplicity, it is convenient to define

\begin{equation}
R_+ = \left( \frac{1}{G_{\uparrow}}+\frac{1}{G_{\downarrow}} \right)
\end{equation}
and

\begin{equation}
R_- = \left( \frac{1}{G_{\uparrow}}-\frac{1}{G_{\downarrow}} \right).
\end{equation}
In this notation the voltage drop at the interface is

\begin{equation}
V = \frac{j}{4}\left( R_+ + R_- P \right)
\end{equation}
where P is the current density spin polarization $P= (2\beta_0 -1)$.  Including the
voltage derivative of the interface conductances gives the voltage signal as

\begin{equation}
\frac {dV}{d\delta^b} = \frac{j}{4}\left[ \left( \frac {d(R_+ + R_- P)}{dV} \right)\frac{dV}{d\delta^b}
+R_- \frac{dP}{d\delta^b} \right]
\end{equation}
where $\left( \frac {d(R_+ + R_- P)}{dV} \right)\frac{dV}{d\delta^b}$ is the new term.
Solving for $\frac{dV}{d\delta^b}$ gives

\begin{equation}
\frac{dV}{d\delta^b} = \frac{\frac{j}{4}R_-
\frac{dP}{d\delta^b}}{1-\frac{j}{4}\left( \frac {d(R_+ + R_- P)}{dV} \right)}.
\label{eq:eqna1}
\end{equation}
Equations (\ref{eq:eqn5d}) and (\ref{eq:eqn6d}) are used to find $\frac{dP}{d\delta^b}$

\begin{equation}
\frac{j}{4} \frac{dP}{d\delta^b} = \frac{\nu}{\Delta}
\label{eq:eqna2}
\end{equation}
where

\begin{equation}
\nu= \frac{2E_F}{3e} \left[ \frac{1}{(1+2 \delta^0)^{1/3}} +
\frac{1}{(1-2 \delta^0)^{1/3}} \right] \frac{\Lambda_-}{\Lambda_+}F(j)
\label{eq:eqna3}
\end{equation}

\begin{eqnarray}
\Delta = R_+ - \frac{2 \Lambda_-}{\sigma}\left[ \frac{1}{(1+2 \delta^0)^{1/3}} +
\frac{1}{(1-2 \delta^0)^{1/3}} \right] + \nonumber\\
\frac{\frac{jR_-}{4}\frac{d \left( R_+ P + R_- \right)}{dV}}{1-\frac{j}{4} \frac {d(R_+ + R_- P)}{dV} }.
\label{eq:eqna4}
\end{eqnarray}
Equations (\ref{eq:eqna1}), (\ref{eq:eqna2}), (\ref{eq:eqna3}), and (\ref{eq:eqna4}) give the voltage signal including
the voltage derivative of the interface conductances.
The new terms that contain $\frac {d(R_+ + R_- P)}{dV}$ or $\frac{d(R_+ P + R_-)}{dV}$ are numerically small for
the parameters considered here.

\bibliography{model-1.7}% Produces the bibliography via BibTeX.

\begin{thebibliography}{24}
\bibitem[{G(V()}]{G(V)}
\bibinfo{note}{Experimentally the change in voltage, at constant current, is
  measured as the spin generation, $C$, is varied. Because the interface
  conductances $G_{\uparrow,\downarrow}$ are voltage dependent and the voltage
  changes with changes in $C$, there are also terms proportional to the voltage
  derivative of the interface conductances. These terms are small for typical
  parameter values. These terms are easily included, but we neglect them here
  for algebraic simplicity and include a discussion of them in the appendix.}
\bibitem[{lar()}]{largebias}
\bibinfo{note}{The linearity condition $2\delta^0 = small$ can be violated at
  large bias if the the interface conductances are strongly spin dependent. The
  asymptotic results described here are qualitative. The numerical results,
  presented in the next section, do not assume the linearity condition.}
\end{thebibliography}

\end{document}